\documentclass{emulateapj}

\shorttitle{Pulsational Instability in Accreting WDs}
\shortauthors{Arras, Townsley, Bildsten}

\begin{document}

\title{Pulsational Instabilities in Accreting White Dwarfs}

\author {Phil Arras$^1$, Dean M. Townsley$^2$, and Lars Bildsten$^{1,3}$}
\affil{$^1$Kavli Institute for Theoretical Physics, Kohn Hall, University of
California, Santa Barbara, CA 93106}

\affil{$^2$Department of Astronomy and Astrophysics,
5640 South Ellis Avenue, University of Chicago, Chicago, IL 60637}

\affil{$^3$Department of Physics,
University of California, Santa Barbara, CA 93106\\
arras@kitp.ucsb.edu; townsley@uchicago.edu; bildsten@kitp.ucsb.edu}

\begin{abstract}

The Cataclysmic Variable (CV) population harbors a diverse range of
donor stars and accreting white dwarfs (WDs). A range of WD masses is
expected, from low mass Helium core WDs, to massive WDs which have
previously accreted at rates high enough for Hydrogen to burn
steadily. Furthermore, a wide range of Helium enrichment is expected
in the accreted material depending on the degree to which the donor
star is evolved. We investigate the impact of this diversity on the
range of effective temperatures ($T_{\rm eff}$) for which g-modes are
unstable. Motivated by earlier theoretical studies, we use a simple
criterion for g-mode excitation: that the thermal time at the base of
the convection zone becomes longer than (some multiple of) a fiducial
shortest g-mode period. The critical $T_{\rm eff}$ below which modes
are unstable (``blue edge") then depends on both surface gravity, $g$,
and He abundance, $Y$. The Hydrogen/first Helium ionization
instability strip is more sensitive to $g$ than $Y$. We find that (for
solar composition envelopes), relative to a fiducial WD mass $0.6\
M_\odot$, the blue edge for a $0.4\ M_\odot$ He core WD shifts
downward by $\approx 1000\ {\rm K}$, while that for a massive $\approx
1.2\ M_\odot$ WD shifts upward by $\approx 2000\ {\rm K}$.
The second Helium ionization instability strip exhibits strong
dependences on both $g$ and $Y$. Surprisingly, increasing $Y$ by only
$10\%$ relative to solar creates an instability strip near $15,000\
{\rm K}$. Hence CV's below the period gap with evolved donor stars of
$Y\ga 0.4$ may have an ``intermediate" instability strip well outside
of the isolated DA and DB variables.  This ``intermediate" instability
strip also occurs for low mass He WD with solar composition envelopes.
The lack of pulsations in CV's with $T_{\rm eff}$ in the pure Hydrogen
ZZ Ceti instability strip is also easily explained.

\end{abstract}

\keywords{binaries: close---gravitational waves--novae, cataclysmic --- variables--- white dwarfs}

\section{Introduction}

As they cool in isolation, WDs cross regions of pulsational instability
and undergo non-radial oscillations, in particular g-modes (see
\citealt{GautSaio96}).  Those with pure H atmospheres (DAV) cross the ZZ
Ceti instability strip when $T_{\rm eff}\approx 11,000-12,000 {\rm K}$
(see Mukadam et al 2004; Gianninas et al. 2005), whereas those with pure He
envelopes (DBV) become unstable when $T_{\rm eff}\approx 22,400-27,800\ {\rm K}$
\citep{Beauetal99}. The composition of these envelopes is rather pure, and
the range of WD masses is that expected from isolated stellar evolution,
favoring $M\approx 0.6M_{\odot}$. Careful analysis of the adiabatic
pulsations for these WDs yield accurate measurements of the WD masses ($M$),
shell masses, and spin (see e.g. \citealt{Brad01}; \citealt{Kepletal03}).

The recent discoveries of similar period pulsations from accreting WDs
in CVs provides a window to their internal properties. The accretion
of $>0.1M_\odot$ over $10^9$ years impacts both the WD's rotation rate
and mass, parameters easily diagnosed through seismic studies.
Starting with the initial discovery \citep{vanZetal00} of 3
oscillation periods in the WD primary of GW Lib, there are now $ 10$
(\citealt{WarnWoud03}; \citealt{WoudWarn04}; \citealt{Arauetal05};
\citealt{Pattetal05}; \citealt{Szkoetal05}; \citealt{Vanletal05};
\citealt{Gansetal06}) pulsating WDs in CVs.  These are all found in
CVs below the period gap,  where a quiescent spectrum shows
evidence for a WD with $T_{\rm eff}<25,000 \ {\rm K}$ \citep{1991AJ....102..295S,1999PASP..111..532S}. Calculations of
WD heating by prolonged accretion \citep{TownBild04} explain the
$T_{\rm eff}$'s at these orbital periods, and were
employed for the first adiabatic analysis of GW Lib, yielding both the
WD and accreted layer mass \citep{Townetal04}.

  We show here that the expected diversity of WD masses and
accreted layer compositions (in particular, enriched He abundance from
an evolved donor) in these systems will lead to a broad range of
$T_{\rm eff}$'s where WDs will pulsate. For example, low-mass He core
WDs can  pulsate at $T_{\rm eff}\approx 10,000 \ {\rm K}$, well
below the ZZ Ceti instability strip. The enhanced Helium abundance
expected from an evolved secondary can also push the instability strip
to values as high as $20,000 \ {\rm K}$. These effects likely explain
why many of the observed pulsators in CVs are ``outside'' of the
conventional ZZ Ceti instability strip appropriate to pure H envelopes
\citep{SzkoGWLib02}.

In \S 2, we clarify the influence of WD mass and envelope composition
on mode driving. Assuming that Brickhill's convective driving
mechanism operates, we explore the depth of the convection zone as a
function of composition and surface gravity. A near-solar mix of
elements allows the possibility of both H/HeI and HeII ionization
zones, opening up additional frequency ranges for instability of 
hotter and/or rapidly rotating WD. Section 3 contains a discussion of
the expectations for the occurrence of He core WDs in CVs, and
explains why many CVs can have an evolved donor 
that provides He rich material to the WD envelope. We close in \S 4 by
summarizing our new understanding and discussing future work.

\section{ Gravity Wave Excitation }

Accreting WDs consist of a geometrically thin, mostly non-degenerate
accreted envelope overlying a degenerate core. The envelope
composition (of mean molecular weight $\mu$) is inherited from the
donor star, while the core composition can be Helium for low mass ($<
0.45\ M_\odot$) WDs or a mixture of Carbon and Oxygen for larger
masses ($ 0.6-1.1 M_\odot$). A luminosity from ``compressional
heating'' of accretion is generated deep in the envelope and core
(Townsley \& Bildsten 2004), hence the luminosity is nearly constant
at the shallow depths where mode driving (and damping) occurs.

The composition discontinuity at the base of the accreted envelope
(where the temperature is $T_b$)
separates the WD into two resonant cavities: the envelope and the core
(Townsley et al. 2004). The highest frequency gravity wave (g-mode)
lives mainly in the envelope (and has no radial nodes there) and 
has a frequency $\omega \sim (gH)^{1/2} k_\perp \sim 2\pi /
(100\ {\rm s})$ (for $l=2$), where $H$ is the scale height at the
base, $(gH)^{1/2}\simeq (k_B T_b/\mu m_p)^{1/2}$ is roughly
the sound speed at the base, and $k_\perp=[l(l+1)]^{1/2}/R$ is the
horizontal wavenumber. Internal gravity waves trapped in the envelope
have frequencies which are smaller by $\approx 1/n$, where $n$ is the
number of radial nodes there. 

 Gravity waves are driven overstable by converting part of the
escaping heat flux into mechanical motion. This occurs near ionization
zones, either due to a rapid outward increase in opacity in radiative
zones (``the $\kappa$ mechanism", \citealt{Dzie77}), or more likely,
due to a convection zone caused by such rapid increase in opacity
(Brickhill's ``convective driving" mechanism, \citealt{Bric83},
\citealt{WuGold99}).  We focus on Brickhill's mechanism, where the 
key quantity is $\omega \tau_{\rm th,bcz}$, the product of the mode
frequency and the thermal time, $\tau_{\rm th,bcz}$, 
at the base of the convection zone. When $\omega \tau_{\rm th,bcz}
\ga 1/4$ \citep{WuGold99}, the thermal adjustment time in the
convection zone is longer than the mode period, and flux perturbations
are absorbed by the convection zone, acting to put energy into the
oscillation. In this paper, we adopt the instability criterion $P \leq
8\pi \tau_{\rm th,bcz}$, where $P=100\ {s}$ is the fiducial shortest
period g-mode. We will not calculate the ``red edge" $T_{\rm eff}$ of
the instability strip, below which the pulsations are
unobservable. For isolated DAVs, the instability strip is quite narrow
($\approx 1000\ {\rm K}$; \citealt{MukaWingetal04}), and is much broader
($\approx  5000\ {\rm K}$) for the DBVs \citep{Beauetal99}. Lastly, when
two convection zones are present, we only consider driving by the
inner convection zone as this most closely resembles the theory
derived for the DAV's (\citealt{Bric83,WuGold99}). However, driving
may also be possible by the outer convection zone, if the radiative
layer between the two convection zones does not contribute
too much damping. We leave this issue for future studies.

\begin{figure*}
\epsscale{1.15}
\plotone{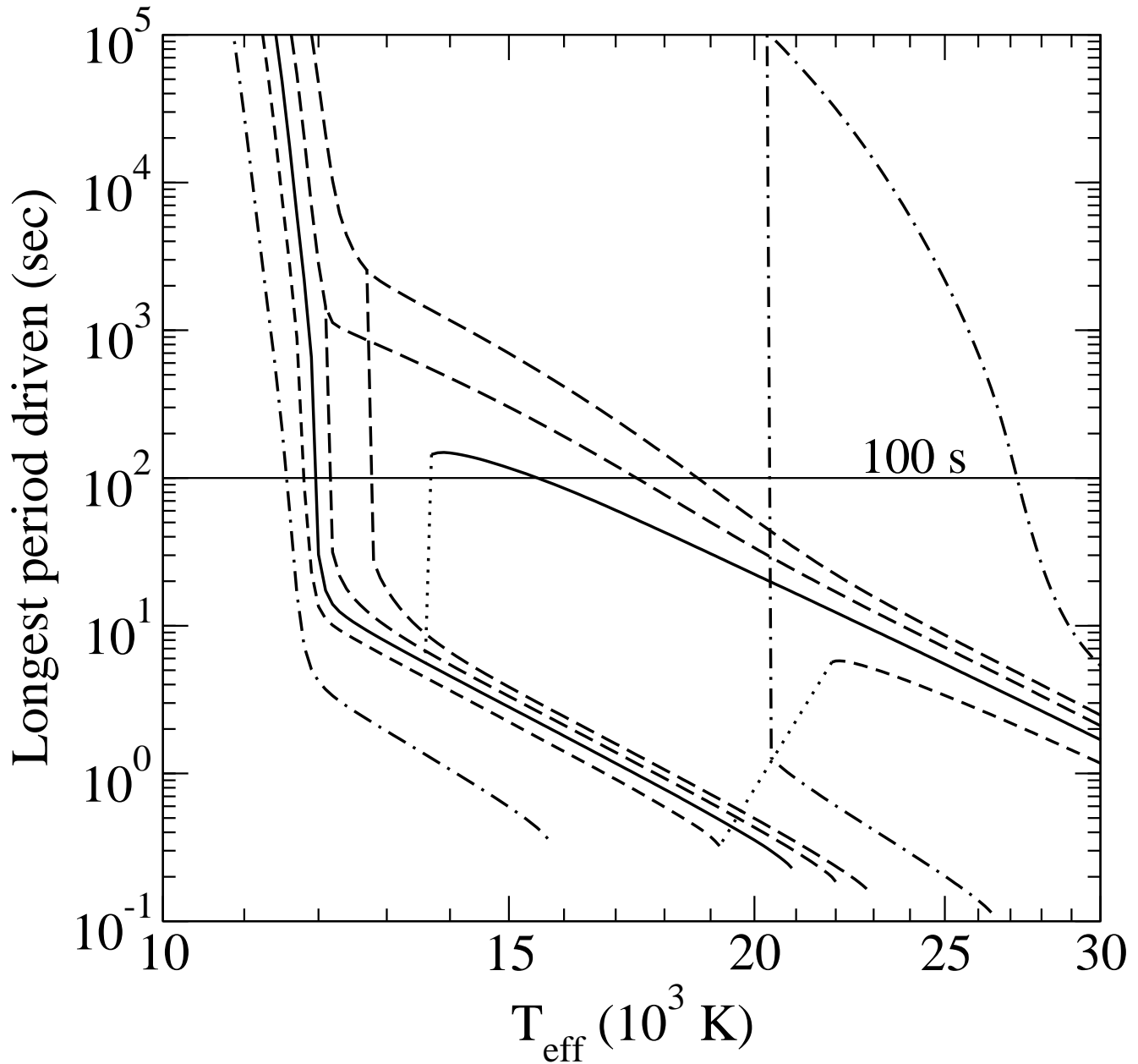}
\caption{ \label{fig:tthbczXH}
Longest period driven mode, $8\pi \tau_{\rm th,bcz}$, vs. $T_{\rm eff}$ for different H
to He ratios, but fixed gravity $g=10^8\ {\rm cm\ s^{-2}}$ and metallicity
$Z=0.02$. A fiducial shortest period g-mode, shown at $100\ {\rm s}$,
is driven when the curve is above the horizontal line.
There can be two convection zones associated with second
He ionization (HeII, upper branches), or first He ionization and/or H
ionization (H/HeI, lower branches).  Lines are shown for $Y=0.0$ (left
dot-dashed), solar composition ($Y=0.28$, leftmost dashed),
$Y=0.38$ (solid), $0.48$ (middle dashed),  $0.58$ (rightmost
dashed), and $0.98$ (right dot-dashed).  Dotted lines only connect related
curves and do not represent convection zone boundaries.
}
\end{figure*}

The variation of $\tau_{\rm th,bcz}$ with $g$ and $Y$ is found by constructing atmosphere
models that use the OPAL opacity and equation of state
\citep{IgleRoge96} with solar metallicity ($Z=0.02$). These are plane
parallel, constant gravity and flux envelopes, using the ML2 mixing
length prescription \citep{Bergetal92} in convective regions, in which
the mixing length is set equal to the pressure scale height.

Figure \ref{fig:tthbczXH} shows the longest driven mode period, $8\pi
\tau_{\rm th,bcz}$, as a function of $T_{\rm eff}$ for $g=10^8\ {\rm
cm\ s^{-2}}$ and a range of $Y$. At a given $T_{\rm eff}$ there can be
two distinct convection zones: an outer convection zone due to H and
first He ionization (H/HeI) at short thermal times, and an inner
convection zone due to second He ionization (HeII) at long thermal
times. The six different lines show $Y=0$ to $Y=0.98$, from left to
right.  For solar abundance ($Y= 0.28$, leftmost dashed line), we find
that $8\pi\tau_{\rm th,bcz}$ exceeds 100 s for $T_{\rm
eff}\lesssim12,000$ K; the blue edge is not raised significantly by
addition of a solar fraction of He relative to a pure H atmosphere. At
solar He fraction, the inner HeII convection zone is only present at
high temperatures ($\gtrsim22,000$ K) and only drives $\ell \gg 1$
modes with periods $P \la 10$ s, which are unobservable.  However,
increasing $Y$ above solar, even by a moderate amount, has a dramatic
effect on the HeII convection zone. For $Y=0.38$ (solid line), the
inner HeII convection zone is present down to $T_{\rm
eff}\simeq14,000$ K, and raises the blue edge to near 16,000 K. We
call this the ``intermediate" instability strip.

The $Y=0.38$ line also indicates a feature of mode driving in CV WDs,
in which a single object can have two instability strips.  The upper
strip in $T_{\rm eff}$, arising from the HeII convection zone, has a red
edge determined by the truncation of the HeII convection zone at lower
$T_{\rm eff}$, rather than Brickhill's (\citeyear{Bric83}) red edge
due to vanishing flux perturbation at the surface.  Instability can
then reappear at $T_{\rm eff}\lesssim 12,000$ K.  This only occurs
over a limited range of $Y$.  At $Y=0.48$ (middle dashed line),
the HeII convection zone is not truncated, and below the blue edge
it extends across $8\pi\tau_{\rm th}=100$ s.  Rather than the H/HeI
convection zone rising up to create another instability strip, the two
convection zones merge when the radiative region between them disappears.
The presence of H  will, however, create an extremely wide instability
strip by keeping the convection zone thin compared to the $Y=0.98$ case
(right dot-dashed line).

\begin{figure*}
\epsscale{1.15}
\plotone{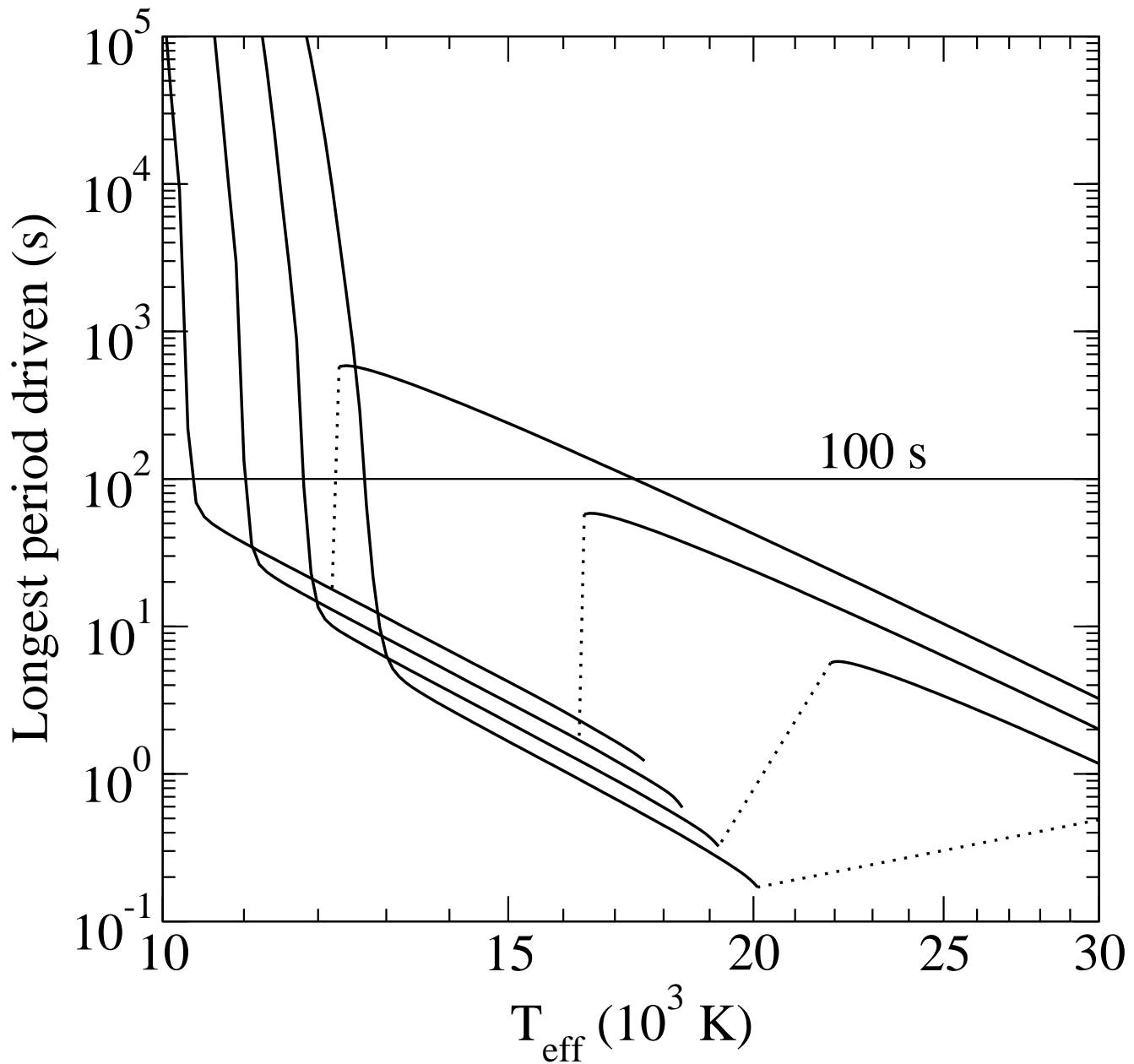}
\caption{ \label{fig:tthbczg}
Longest period driven mode, $8\pi\tau_{\rm th,bcz}$, vs. $T_{\rm eff}$ for
different gravity, but fixed (solar) composition $Y=0.28$ and $Z=0.02$.  The
gravity for each line is, from left to right at low $T_{\rm eff}$,
$g=10^{7.0}$, $10^{7.5}$, $10^{8.0}$, and $10^{8.5}$ cm s$^{-2}$.  Dotted
lines only connect related curves and do not represent convection zone
boundaries.
}
\end{figure*}

\begin{figure*}
\epsscale{1.15}
\plotone{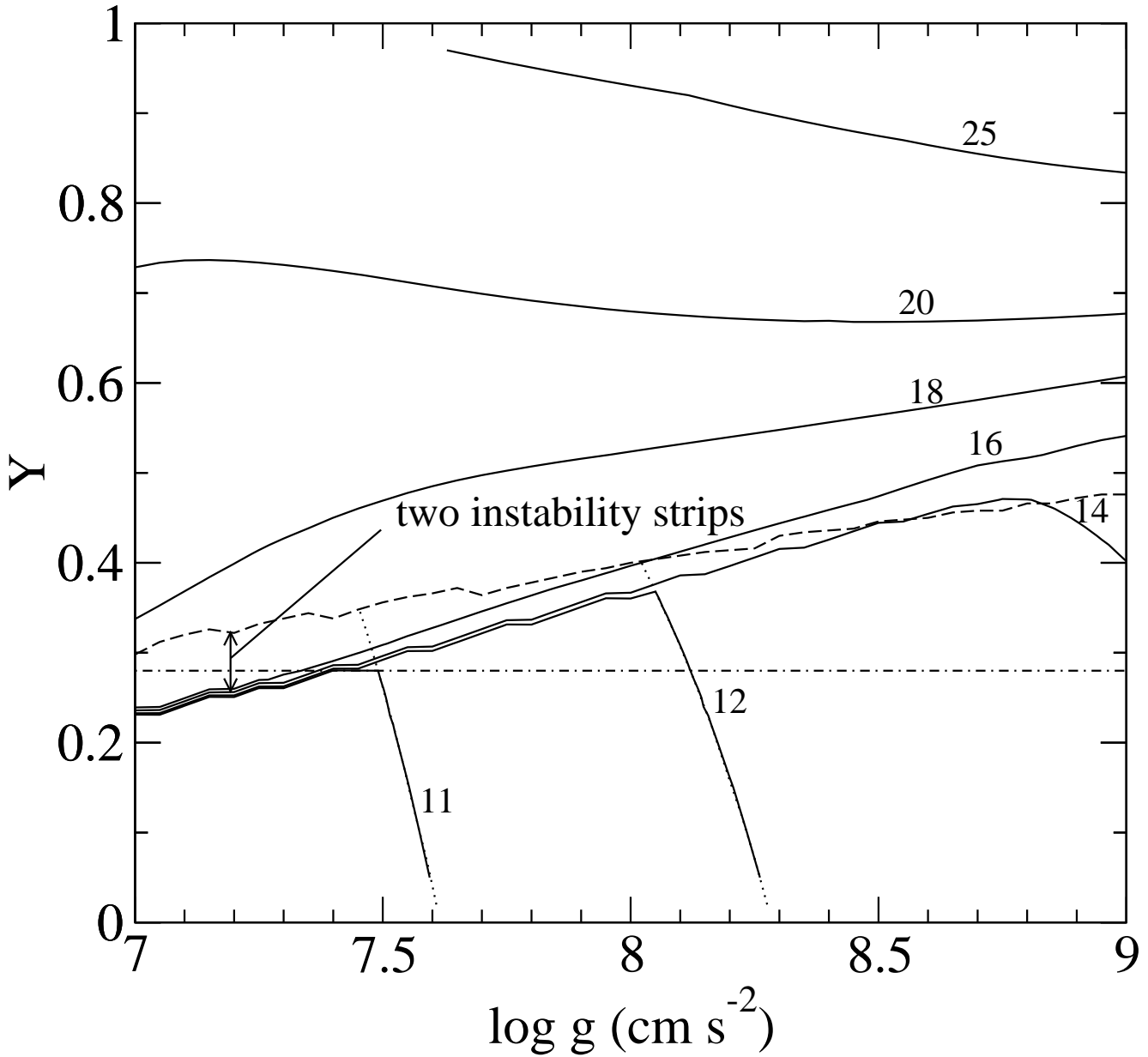}
\caption{\label{fig:blueedge}
Contours of constant $T_{\rm eff}$ for the blue edge of the instability
strip in gravity, $\log g$, and Helium mass fraction, $Y$.  The blue
edge is approximated by the $T_{\rm eff}$ for which $8\pi \tau_{\rm th,bcz}
=100$ s.  For large $Y$, driving is due to the HeII convection zone,
while for small $Y$ it is due to the H/HeI convection zone. The dashed
line indicates the $Y$ above which the HeII convection zone is always
present. The sudden increase of the blue edge with $Y$ occurs where the
HeII convection zone reaches above 100 s.  Between this and the dashed
line, there will be two separate instability strips caused by the absence
of the HeII convection zone at some $T_{\rm eff}$'s.  The blue edge of the
lower instability strip is shown by the dotted lines.  Solar composition,
$Y=0.28$, is indicated by the dash-dotted line.  } 
\end{figure*}

A similar experiment is shown in Figure \ref{fig:tthbczg}, where
$g$ varies from $g=10^{7.0}$ to $10^{8.5}$ cm s$^{-2}$ keeping
solar composition ($X=0.7$, $Y=0.28$, $Z=0.02$).  A gravity of
$g=10^{7.5}$ cm s$^{-2}$ corresponds to a $M\approx 0.4 M_\odot$ WD,
while $g=10^8$ cm s$^{-2}$ gives $M \approx 0.6 M_\odot$, and $g=10^9$
cm s$^{-2}$ gives $M \approx 1.2 M_\odot$. As in Figure
\ref{fig:tthbczXH}, the blue edge is the intersection of a curve with
the horizontal line representing the fiducial shortest period
g-mode. As discussed by \citet{ArraTownBild05}, the blue edge for the
H/HeI ionization zone moves up (down) by $\approx 2000\ {\rm K}$ for a
factor of $10$ increase (decrease) in gravity. Hence, a blue edge
range of $\approx 4000\ {\rm K}$ is expected from the lowest mass
He-core WD to the massive $\ga 1.1 M_\odot$ Carbon/Oxygen WD. In
short, the range of WD masses expected in CVs should extend the H/HeI
instability strip to be both well above and below the isolated WD ZZ Ceti
instability strip.  For the HeII ionization zone, note the appearance
of the ``intermediate" instability strip at $\approx 15,000\ {\rm K}$
for low mass WDs. Since $\tau_{\rm th,bcz}$ decreases as $g$ increases,
the blue edge for HeII ionization decreases with higher $g$. This behavior
is opposite that found in the nearly pure He case, where $\tau_{\rm th,bcz}$
increases as $g$ increases (see figure \ref{fig:blueedge}).

To elucidate the sensitivity of the blue edge to the wide range of
gravities and enrichments possible for CV WDs, Figure \ref{fig:blueedge}
shows contours of constant blue edge $T_{\rm eff}$ in $\log g$ and $Y$
space.  At small $Y$, driving occurs due to the H/HeI convection zone,
while at large $Y$ driving is due to the HeII convection zone.  For low
and moderate $g$, the transition occurs abruptly when the HeII convection
zone extends above 100 s, causing the blue edge to jump up by several
thousand K.  The dashed line indicates the $Y$ above which the HeII
convection zone is always present. As discussed above, there is a small
region, indicated on the plot, where there are two separate instability
strips.  Dotted lines show the blue edge of the lower instability strip.
We find that starting from solar abundance ($Y=0.28$, dot-dashed line),
$g$ must be increased significantly, half an order of magnitude, in
order to increase the blue edge $T_{\rm eff}$ by 1000 K. In contrast, a
modest increase in $Y$ can effect a very sharp increase in the blue edge.
There are thus two types of accreting pulsators, those with blue edge
$T_{\rm eff}\ga 15,000\ {\rm K}$ that are slightly enriched, and those
closer to solar composition with blue edge $T_{\rm eff}\la 15,000\
{\rm K}$, more similar to normal ZZ Ceti stars. Observationally,
these two instability strips may overlap in $T_{\rm eff}$ depending on
the location of the red edge.

\section{Helium Enrichment in Evolved Cataclysmic Variables}

 Both theory and observation (see Townsley \& Bildsten 2003) tell us
that non-magnetic CVs need to be below the period gap ($P_{\rm orb}\le
2 \ {\rm hr}$) to reach the $T_{\rm eff}<25,000 {\rm K}$ range where
pulsations are possible. This is indeed where the known pulsating WDs
in CVs occur. We showed in the previous section that, due to the
dependence on $Y$ and $g$, the instability strips for accreting WDs
will be richer than in isolated DA or DB WDs.

   The first diversity to consider is the WD mass.  The expectation
(see Howell, Nelson \& Rappaport 2001; hereafter HNR) is that WDs
below the period gap will have a large range of masses, all the way
down to low-mass He core WDs ($M<0.45 M_\odot$). HNR's population
synthesis calculations showed that as many as $20\%$ of the CVs 
with $P_{\rm orb}<2 \ {\rm hr}$ will have low mass He WDs. These
will have surface gravities much lower than any of the isolated DAVs,
for which the recent tabulations (Mukadam et al 2004; Gianninis et
al. 2005), show no DAV with $\log g\le 7.7-7.8$. The pulsator HS
2331+3905 has a low $T_{\rm eff}\simeq 10,500$ K \citep{Arauetal05},
which from Figure \ref{fig:blueedge} is quite reasonably inside the
instability strip for solar abundance with $\log g \lesssim7.5$.  The
inferred low mass of 0.4 $M_\odot$ makes HS 2331+3905 a candidate He
core WD. 

The instability strip is also sensitive to the He abundance in the WD
envelope. At accretion rates $\gtrsim 10^{-14}$ $M_\odot$ yr$^{-1}$
for $T_{\rm eff}\gtrsim 11,000$ K, there is no time for relative H/He
separation \citep{Paquetal86}. Since even the small amount of
quiescent disk accretion exceeds this rate, the envelope He abundance
is set by the donor star. If the donors were unevolved, then
the range would be narrow, $Y=0.25-0.3$, only depending on the initial
metallicity of the donor. However, Pylyser \& Savonije (1989) raised
the distinct possibility of initial donor masses for CVs with $M\ge
1M_\odot$, which, depending on the age of the system at the onset of
mass transfer, can have undergone nuclear burning in the core to raise
the He abundance. Such donors are required to explain
supersoft sources as stably burning WDs (van den Heuvel et al. 2002).
By the time such donors are below the period gap and fully convective,
such a core He enrichment will be evident at the surface and in the
accreting matter. This gives reason to explore the range of 
$Y=0.3-1.0$.

 The prevalence of evolved donors amongst CVs with $P_{\rm orb}<8 $
hours depends on many factors, including the efficiency of the common
envelope phase, the outcome of the rapid thermal timescale mass
transfer expected at the onset of Roche lobe filling, and the nature
of magnetic braking in the pre-CV stage.  (Schenker et al. 2002;
Podsiadlowski, Han \& Rappaport 2003; Andronov \& Pinsonneault 2004;
Kolb \& Willems 2005). Podsiadlowski, Han \& Rappaport (2003)
performed a binary population synthesis calculation that allowed for
evolved donors with a range of initial He abundances amongst initially
massive ($M>M_\odot$) donors. They found that these could actually be
the dominant population for $P_{\rm orb}>5$hours, but that the large
number of CVs injected at shorter orbital periods dilutes their
fraction at shorter periods. Below the period gap, between one in ten
and one in three would have evolved companions with $Y>0.5$ The previously
discussed observables for evolved donors are a later spectral type (at a
fixed orbital period), and exposed material that is Nitrogen enhanced and
Carbon poor due to the CN cycle. This latter hypothesis has been tested
by G\"ansicke et al. (2003) through measurements of UV line ratios in a
diverse set of CVs, both above and below the period gap.  They concluded
that as many as 10-15$\%$ of their sample could have evolved donors.

We are introducing here an additional probe of evolution, which is the
impact of the He abundance on the instability strip. More evolved
donors will show pulsators at hotter $T_{\rm eff}$ than those which
are unevolved. In the absence of more detailed population study
predictions of the He abundance (typically calculated, but not
reported in depth since, until now, no observational probe of He
abundance was available), we can say very little. Due to mass transfer
stability considerations, the donor stars for a He core WD should be
initially low mass and therefore unevolved. Hence the correct low
gravity models to consider are those with $Y=0.25-0.3$, but no
higher. In the opposite sense, we would expect highly evolved donors
to have preferentially more massive WDs.  Whether these correlations
lead to a range of $T_{\rm eff}$'s or other correlations remains to be
seen.

\nocite{ Scheetal02, PylySavo89, Podsetal03, KolbWill05, Howeetal01,
Gansetal03, AndrPins04, Gianetal05, Heuvetal92}

\section{Conclusions and Future Work}
\label{conclusion}

We have shown that detecting pulsations in accreting WDs below the
period gap has the promise to reveal the diversity expected in both
the WD masses and the accreted He abundance. We do not expect one
instability strip as known for isolated DA WDs, but rather a revealing
diversity of $T_{\rm eff}$'s.  In particular, hot CV primaries like
that in GW Lib with $T_{\rm eff}\simeq14,000$ K can pulsate due to
enhanced Helium abundance, or extreme surface gravity (large or
small). Cool CV primaries like that in HS 2331+3905 with $T_{\rm
eff}=10,500$ K, can pulsate due to low surface gravity.  Both of these
possibilities are naturally present in the diversity expected within
the population of CVs.  Indeed, some systems can have two separate
instability regions in $T_{\rm eff}$.

Little is known with certainty about the WD mass distribution, but
evolutionary models provide a reasonable expectation for the level of Helium
enrichment of the accreted material.  CVs in which the companion is the
remaining core of a slightly evolved star can have He significantly enriched
above solar.  Below the period gap, between one in ten and one in three CVs
should have evolved companions with $Y>0.5$ \citep{Scheetal02,Podsetal03}.

Important work which remains is validation of our assumption that the
numerical factor relating $\tau_{\rm th,bcz}$ and the driven mode
period does not vary significantly with either $g$ or $Y$. A large
variation (more than unity) is not expected, but quantifying this
statement is necessary for using the presence or absence of modes to
constrain properties of particular systems. Additional damping due to
the opacity effects of metals in the radiation zone may affect this
prefactor, shifting the blue edges evaluated here to lower $T_{\rm
eff}$. Inefficient convection in surface layers would decrease the rate
of convective driving, again causing the blue edge to shift downward.

The WDs in CVs are expected to be rotating rapidly due to the angular
momentum gain from long term disk accretion. This can qualitatively
change the frequency spectrum of non-radial oscillations.
\citet{Dzie77} showed that $g$-modes with large numbers of angular
nodes, and hence large frequency, can be driven even in relatively hot
WDs by the $\kappa$-mechanism acting in the He-II ionization
zone. Such modes are unobservable in non-rotating stars due to
averaging of the flux perturbation over the stellar disk. Rotationally
modified g-modes, which are squeezed into an equatorial band, may be
both unstable and observable, since they can have large horizontal
wavenumber (and frequency) for a small number of angular nodes.

\acknowledgments

We thank Peter Goldreich for a useful discussion. We would
especially like to thank Boris G\"ansicke and the anonymous referee for
their careful reading and insightful comments which improved this paper.
This work was supported by the National Science Foundation (NSF) under
grants AST02-05956 and PHY99-07949. D.M.T. is supported by the NSF
Physics Frontier Centers' Joint Institute for Nuclear Astrophysics under
grant PHY02-16783 and DOE under grant DE-FG 02-91ER 40606. Support for
this work was provided by NASA through grant GO-10233.03-A from STScI,
which is operated by AURA, Inc under NASA contract NAS 5-26555


\end{document}